\documentstyle[preprint,aps]{revtex}

\preprint{\bf PREPRINT}

\begin{document}
\columnsep0.1truecm
\draft
\title{Avalanches, Barkhausen Noise, and Plain Old Criticality}
\author{Olga Perkovi\'c, Karin Dahmen, and James P. Sethna}
\address{Laboratory of Atomic and Solid State Physics,\\
Cornell University, Ithaca, NY, 14853-2501}
\maketitle

\begin{abstract}
We explain Barkhausen noise in magnetic systems in terms of avalanches
near a plain old critical point in the hysteretic zero-temperature
random-field Ising model. The avalanche size distribution has a
universal scaling function, making non-trivial predictions of the shape
of the distribution up to 50\% above the critical point, where two
decades of scaling are still observed. We simulate systems with up to
$1000^3$ domains, extract critical exponents in 2, 3, 4, and 5
dimensions, compare with our 2d and $6-\epsilon$ predictions, and
compare to a variety of experimental Barkhausen measurements.

\end{abstract}


\narrowtext

When materials are pushed, they can yield in different ways.  Some
crackle: they transform through a series of pulses or avalanches.  In
many systems, the behavior of these avalanches is unaffected by thermal
fluctuations: one domain triggers, pushing some of its neighbors to trigger,
in a deterministic process dependent on the static, quenched disorder
in the material (and on the stress history).  The statistics of Barkhausen
noise (the avalanches seen in magnetic materials as the external
magnetic field is ramped up and down) has been extensively studied
experimentally \cite{Barkhausen-review,Cote,OBrien-Weissman,Stierstadt,%
Bertotti94,Bittel,Lieneweg,Lieneweg-and-Grosse-Nobis,Bertotti90,Urbach,%
Montalenti}.
We suggest that the zero-temperature random-field Ising
model\cite{hysterI,hysterII} provides a universal, quantitative explanation
for many of these experiments.

A typical experiment will collect a histogram of pulse sizes, times, or
energies.  The distribution will follow a power law, which cuts off
after two to several decades --- much broader than any observed
morphological feature in the materials.  An explanation for the
experiment must involve collective motion of many domains; it must
provide an explanation for the observed power-law scaling regions, {\it
and it must provide an explanation for the cutoff}.

Figure~1 shows the distribution $D_{\it int}(S, R)$ of avalanche sizes
for our model in 3d (discussed already in reference \cite{hysterI}), at
several values of the microscopic disorder $R$.  The
model is a collection of domains $s_i=\pm 1$ coupled to an external
field $H$, a local random field $f_i$ chosen from a distribution
$\rho(f) = \exp(-f^2/2 R^2)/ \sqrt{2 \pi} R$ of standard deviation $R$,
and to its nearest neighbors $s_j$ with an energy of strength $J=1$.
The domain $s_i$ flips over when the net local field $F_i \equiv f_i + H
+ J \sum_{nn} s_j$ seen at site $i$ changes sign.  Due to the
nearest-neighbor interaction, a flipping spin often causes one or more
neighbors to flip also, thereby spawning a whole avalanche of spin
flips.  Figure~1 shows the avalanches found by integrating as
the external field $H(t)$ is raised adiabatically from $-\infty$ to $\infty$
(the field is thus constant during the individual avalanches).

Notice three things about figure~1.  (1)~The distributions follows
a power law, which cuts off after two to several decades.  (2)~The
cutoff appears to diverge at a critical value of the disorder
$R_c$, which we estimate in three dimensions to be $2.16 J$.  (3)~The
critical region is large!  While the true power-law
distribution is only obtained at $R_c = 2.16$, we get avalanches with more
than a hundred domains all the way up at $R=4$.  This suggests that
experiments can see decades of scaling without working hard to find
the critical disorder.  Several decades of scaling without tuning
a parameter need not be self-organized criticality: it can be
vague proximity to a plain old critical point.

Notice four more things about figure~1.  (4)~The straight line lying
askew below the numerical data is our prediction for the asymptotic power
law $D_{\it int}(S, R_c) \sim s^{-(\tau + \sigma \beta \delta)}$.
The obvious experimental method of taking the slope on the log-log
plot gives the wrong answer until many, many decades of scaling are obtained!
(5)~The inset shows the collapse of the data onto a
scaling function
\begin{equation}
{\cal D}_{\it int}(s^\sigma (R-R_c)/R) =
	\lim_{R\to R_c} s^{\tau + \sigma \beta \delta} D_{\it int}(S, R)
\label{curlyD}
\end{equation}
which is a universal prediction of our model: real experiments
rescaled in the same way should look the same (apart from overall
vertical and horizontal scales).  This scaling function is quite unusual:
it grows by a factor of over ten before cutting off.  This bump is the
reason that the experiments take so long to converge to the asymptotic
power law.  To make a definite prediction, we have phenomenologically fit
our curve
\begin{eqnarray}
{\cal D}_{\it int}(x) =
	(0.021 &+& 0.002 x + 0.531 x^2 - 0.266 x^3 \\
\label{fitD}
		 &+& 0.261 x^4) \exp(-0.789 x^{1/\sigma}) \nonumber
\end{eqnarray}
where we guess the error in the curve to be less than 10\% within the range
$0.2<x<1.2$.  (6)~In the main figure, the scaling form passes through
our data quite well, even far from $R_c$.  The scaling theory is predictive
for curves with only two or three decades of scaling.  The critical region
starts when the correlation length (and hence the avalanche size cutoff)
becomes large --- not only when the pure power law takes over.  Using
equations (\ref{curlyD}) and (2) and the values
$\sigma = 0.24\pm 0.02$ and $\tau + \sigma\beta\delta = 2.03\pm 0.03$,
an experimentalist should be able to fit any single
histogram of avalanches, shifting only
the overall vertical and horizontal scale factors.  (7)~Widom scaling
(equation (\ref{curlyD}))
forms a powerful tool: only by varying $R$ were we able to extract the
correct critical behavior.  We suggest that
experimentalists try varying some parameter of their system (annealing time
or temperature, grain size, impurity concentration, ...) and observe
the resulting cutoff dependence.  Any family of curves thus generated
should, near the critical point, be fit with three parameters (including
$R_c$ in (1) and (2)).

A comparison of our predicted exponents with power laws
extracted from a number of experiments on magnetic Barkhausen noise in
bulk three-dimensional systems is shown in Table~1.  One of the
experiments\cite{Lieneweg-and-Grosse-Nobis} varied the annealing time,
and saw a shift in the cutoff, but did not extract a critical annealing
time or do scaling collapses.  The range of values for
$\tau+\sigma\beta\delta$ observed in the experiments is compatible with
the range of log-log slopes we observe due to the unusual scaling form
for the integrated avalanche size distribution $D_{\it int}$ discussed
above.

The largest set of experiments measure the avalanche size distribution
in a narrow range of fields ({\it i.e.}, without averaging over the
entire loop): their power laws fits are a measure of the critical
exponent $\tau$.  Integrating over the hysteresis loop changes the power
law, because only near a critical value $H_c$ of the external field do
large avalanches occur. The cutoff in the avalanche size at $R_c$ goes
as $|H-H_c|^{1/\sigma \beta \delta}$: $\sigma$ as discussed above gives
the cutoff dependence with $R-R_c$, and $\beta$ and $1/\delta$, as
usual\cite{hysterI}, give the singularities of the magnetization with
$R-R_c$ and $H-H_c$ respectively (the exponent for any quantity varying
$H$ at $R_c$ is given by multiplying $1/\beta \delta$ by the exponent for
the singularity varying $R$ at $H_c$).  The scaling function
for the non-integrated avalanche size distribution\cite{hysterI},
we find, does not have a bump.  The experimental measurements for $\tau$
are close to our numerical estimate.

The other experiments (pulse durations, power spectra, and pulse
energies) introduce a new exponent combination $\sigma \nu z$. The
correlation length exponent $\nu$ gives the divergence of the
characteristic spatial extent of avalanches with $R-R_c$, and $\nu z$
gives the divergence of the avalanche durations with $R-R_c$.  The
critical exponent $z$ occasionally can depend on the details of the spin
dynamics\cite{NarayanMiddleton}; it is not even clear whether our
simulation must have the same value of $z$ as our
$\epsilon$-expansion.  Nonetheless, the agreement between our
predictions and the measured values are about as good as the agreement
between the different measurements.

We have also investigated the application of our model to other systems.
Many experiments are done in effectively two-dimensional systems
(magnetic hysteresis\cite{Berger}, avalanches as the field is swept
in superconductors\cite{Stuart-Field-Adams}, and avalanches as helium is
injected into Nuclepore\cite{Hallock}); our 2d explorations are still rather
preliminary.  Our model does not fit the avalanche size distributions
measured in 3d martensitic transitions as the temperature is
ramped\cite{Ortin-martensite-aval}: their measurement of the pulse duration
exponent $(\tau + \sigma\beta\delta -1)/\sigma\nu z + 1 \sim 1.6$
is significantly different from our prediction of
$2.81\pm0.11$.  We expect that the long-range anisotropic elastic fields
in martensites likely change the universality class; similarly
the long-range antiferromagnetic demagnetization fields could affect
experiments in certain geometries (see \cite{Urbach}).  Full
explanations about the various exponent combinations measured in different
systems\cite{DahmenII} and detailed discussions of experiments
and systems\cite{ExperimentPaper} are forthcoming.

Figure~2 shows the results for five of our exponents
($\tau+\sigma\beta\delta$, $\tau$, $1/\nu$, $\sigma\nu$, and $\sigma\nu
z$), measured in 2, 3, 4, and 5 dimensions.  (From these one can get
$\beta$ and $\delta$ separately using the exponent
equality\cite{DahmenII} $\tau + \sigma \beta \delta = 2 + \sigma \beta$.)
We measure the exponents in the (unphysical) dimensions four and five in
order to test our renormalization-group
predictions\cite{hysterII,DahmenII} for the behavior near six
dimensions.  First, notice that the numerical values converge nicely to
the mean-field predictions as the dimension approaches six, and
that the predictions of our $6-\epsilon$ expansion do remarkably well.
(The primary role of the renormalization-group treatment, of course, is to
explain why scaling and universality might be expected in these systems.)
The predictions for $1/\nu$ are to fifth order in $\epsilon$: by mapping
our model to all orders\cite{DahmenII} onto the regular Ising model in
two lower dimensions\cite{oopsI}, we have been able to
use\cite{LeGuillou-Kleinert} the series known to order
$\epsilon^5$ for $\nu$.  The other exponents shown have no equivalent in
the equilibrium model: we have developed\cite{DahmenII} a new method for
calculating these avalanche exponents using two replicas of the system.
The dashed lines show a Borel resummation\cite{LeGuillou-Kleinert} of
the series for $1/\nu$, and the predictions to first order for the other
variables.

Second, notice the exponents in two dimensions.  We here conjecture
that the 2d exponents will be $\tau+\sigma\beta\delta = 2$,
$\tau = 3/2$, $1/\nu = 0$, and $\sigma\nu = 1/2$.
It is likely that two is the ``lower critical dimension'' for our system,
below which all avalanches will be finite except at zero disorder.
At the lower critical dimension, the critical exponents are often
ratios of small integers, and it is often possible to derive exact solutions.
For us, using the fact that there can be at most one system-spanning
avalanche in two dimensions, one can derive a special exponent relation
$1/\sigma\nu = d - \beta/\nu = 2 - \beta/\nu$: this ``hyperscaling''
relation is false in mean-field theory, and definitely ruled out numerically
in four and five dimensions, and probably in three\cite{Comment,Perkovic}.
Folklore in the field\cite{Krey} give us two other likely 2d relations:
one each for the exponent giving the decay in space of the cluster
correlation function $\bar\eta = \beta/\nu+4-d =?\, 4-d$
and of the avalanche correlation function
$\eta = 2 + \beta/\nu - \beta\delta/\nu =?\, 1$.  These relations hold
in the lower critical dimension for the Ising model, the Heisenberg model,
and the equilibrium, thermal random-field Ising model.  The first
of these relations is equivalent to the statement
that the avalanches are compact ($1/\sigma\nu = d = 2$).

We must mention that our firm conjectures about the exponents
in two dimensions must be contrasted with our lack of knowledge about the
proper scaling forms.  At the lower critical dimension, the correlation
length typically diverges exponentially as one approaches the critical
point ($\nu \to \infty$).  Some combinations of critical exponents stay
finite (hence $\sigma\nu = 1/2$), but those which diverge and those which
go to zero usually must be replaced by exponents and logs, respectively.
We have used three different RG-scaling {\it ans\"atze} to model the data in
two dimensions.  (1)~We used the traditional scaling form $\xi \sim
(R-R_c)^{-\nu}$, deriving $\nu = 5.3\pm 1.4$ and $R_c = 0.54\pm 0.04$.
These collapses worked as well as any, but the standard form has more free
parameters to fit with.  Also, the large value for $\nu$ (and larger
values still for $1/\sigma = 10\pm 2$) makes one suspicious.
(2)~We used a scaling form suggested by Bray and Moore\cite{BrayMoore} in the
context of the equilibrium thermal random-field Ising model where
$R_c=0$: if they assume that $R$ is a marginal direction, then by
symmetry the flows must start with $R^3$, leading to
$\xi \sim \exp(A/(R-R_c)^2) \equiv \exp(A/R^2)$.  This form
had the fewest free parameters, and most of the collapses were about
as good as the others (except notably for the finite-size scaling
of the moments of the avalanche size distribution, which did not
collapse well once spanning avalanches became common).  (3)~We
developed another possible scaling form, based on a finite $R_c$
and $R$ marginal, which generically has a quadratic flow under
coarse-graining: here $\xi \sim \exp(A/(R-R_c)$.  Here again the
moments did not scale well; we find $R_c=0.54\pm0.04$, quite compatible
with the traditional scaling collapse.  This is not a surprise:
it is always hard to distinguish large power laws
from exponentials.  Amazingly enough, the exponents plotted in figure~2
were largely independent of which scaling form we used!  The error bars
shown span all three {\it ans\"atze}, and are compatible with our
conjectures above.

We are not the only ones to model avalanche behavior in disordered
magnets. There has been much work on depinning transitions and the
motion of individual interfaces\cite{Depinning,SameDepinning}; our
system, with many interacting interfaces, perversely seems much simpler
to analyze.  Many have studied related models with random
bonds\cite{RandomBonds,Vives} and random anisotropies;
random fields are actually rather rare in experimental systems.  We now
believe on symmetry grounds that all these systems are in the same
universality class (as argued numerically\cite{Vives} and previously
shown for depinning\cite{SameDepinning}).  The external field $H_c$ at
the critical point breaks the rotational and up-down symmetries of these
models (and of the experiments!), and the spins which flip far from the
critical point (roughly $M(H_c)$) act as random fields. On the other
hand, we ignore long-range forces (discussed above) and long-range
correlations in the disorder (e.g., dislocation lines and grain
boundaries): these likely will lead to closely related but distinct
universality classes.

We acknowledge the support of DOE Grant \#DE-FG02-88-ER45364 and NSF
Grant \#DMR-9118065. We would like to thank Bruce W. Roberts, Sivan
Kartha, Eugene Kolomeisky, James A. Krumhansl, Mark Newman,
Jean Souletie and Uwe T\"auber for helpful conversations. This work was
conducted on the SP1 and the SP2 at the Cornell National Supercomputer
Facility, funded in part by NSF, by NY State, and by IBM.
Further pedagogical information using Mosaic is available at
http://www.lassp.cornell.edu/sethna/hysteresis/hysteresis.html .

\begin{table}
\begin{tabular}{crl}
\hline
exponents&
\multicolumn{1}{c}{simulation} &
\multicolumn{1}{c}{experiments}\\
&\multicolumn{1}{c}{in 3 dim.} &
\multicolumn{1}{c}{in 3 dimensions}  \\ \hline
$\tau$ &  $1.6\pm0.06$ & 1.74,1.78,1.88\cite{Cote}; \\
& & $1.5 \pm 0.5$\cite{Stierstadt}; \\
& & $1.33$\cite{Urbach}; \\
& & $1.5 -1.7$ \cite{Bertotti94}  \\
$\tau+\sigma\beta\delta$&$2.03\pm0.03$ &
1.73-2.1\cite{Lieneweg-and-Grosse-Nobis} \\
$(\tau-1)/\sigma\nu z +1$&$2.05\pm0.12$
&1.64,2.1,1.82\cite{Cote}; \\
& & 1.7-2\cite{Bertotti94} \\
$(\tau+\sigma\beta\delta-1)/\sigma\nu z +1$&$2.81\pm0.11$&
2.28\cite{Lieneweg-and-Grosse-Nobis} \\
$(3-\tau)/ \sigma\nu z$&$2.46\pm0.17$& around
2\cite{Cote,Bertotti90}\\
$(3-(\tau+\sigma\beta\delta))/
\sigma\nu z$&$1.70\pm0.10$&1.6\cite{Bittel,Lieneweg};
1.8\cite{Montalenti}\\
$(\tau-1)/( 2-\sigma\nu z)+1$&$1.42\pm0.04$&
1.44,1.58,1.60\cite{Cote} \\
\hline
\end{tabular}
\vspace{1cm}
\caption
{{\bf Critical exponents obtained from numerical
simulations\protect{\cite{hysterI,Perkovic}}
and experiments} on Barkhausen noise in different magnetic
materials (Fe, alumel, metglass \protect{\cite{Cote},}
NiS \protect{\cite{Stierstadt},} SiFe
\protect{\cite{Bertotti94,Bittel,Lieneweg},} 81$\%$NiFe
\protect{\cite {Lieneweg-and-Grosse-Nobis},}
AlSiFe \protect{\cite {Bertotti90},} and FeNiCo \protect{\cite
{Urbach}}).
The sample shapes were mostly wires. The quoted exponents were
experimentally obtained
from the pulse-area distribution in a small
bin of the magnetic field $H$ (exponent $\tau$), the pulse-area
distribution integrated over the entire hysteresis loop $(\tau +
\sigma \beta \delta)$, the distribution of pulse durations in a
small
bin of $H$ ($(\tau-1)/\sigma\nu z + 1$), the distribution of
pulse durations integrated over the loop ($(\tau + \sigma\beta\delta
-1)/\sigma \nu z + 1$), the power spectrum of the pulses
in a small bin of $H$ ($(3-\tau)/\sigma\nu z$), the power spectrum
of the pulses integrated over the hysteresis loop
($(3-(\tau+\sigma\beta\delta))/\sigma\nu z$), and the distribution of
pulse energies in a small bin of $H$ ($(\tau-1)/(2-\sigma\nu z) + 1$).
Notice that these experiments are mostly done in geometries which minimize
the effects of demagnetization fields (deadly to our model).
\label{tab:experiments}}
\end{table}

\begin{figure}
\caption{Avalanche size distribution curves in $3$ dimensions integrated
over the external field. From left to right, the first three curves are
for system size $320^3$ and disorders $4.0$, $3.2$, and $2.6$. They are
averages over different initial random field configurations. The last
curve is a $1000^3$ run at $R=2.25$, where $R_c=2.16$ and $r=(R-R_c)/R$.
The inset shows
the scaling collapse of curves in $3$d. The disorders range between
$2.25$ and $3.2$; the top curves at $R=3.2$ show noticeable 10\%
corrections to scaling. In the main figure, the avalanche size
distribution curves obtained from the fit to this data (thin lines) are
plotted alongside the raw data (thick lines). Notice that the scaling
theory is predictive up to $R=3.2$, $50\%$ above $R_c$.  The long-dashed
straight line is the expected asymptotic power-law behavior $s^{-2.03}$.
Notice that it does not agree with the measured slope of the raw data.
\label{avalanches}}
\smallskip
\caption{Numerical values (filled symbols) of the exponents $\tau +
\sigma\beta\delta$, $\tau$, $1/\nu$, $\sigma\nu z$, and $\sigma\nu$
(circles, diamond, triangles up, squares, and triangle left) in 2, 3, 4,
and 5 dimensions. The empty symbols are values for these exponents in
mean field (dimension 6). Note that the value of $\tau$ in $2$d was not
measured. The empty diamond represents the conjectured value (see text). We
have simulated sizes up to $7000^2$, $1000^3$, $80^4$, and $50^5$, where
for $320^3$ for example, more than $700$ different random field
configurations were measured. The long-dashed lines are the $\epsilon$
expansions to first order for the exponents $\tau + \sigma\beta\delta$,
$\tau$, $\sigma\nu z$, and $\sigma\nu$. The short-dashed line is the
Borel sum \protect{\cite{LeGuillou-Kleinert}} for $1/\nu$ to fifth order
in $\epsilon$.  The error bars denote systematic errors in
finding the exponents from collapses of curves at different values of
disorder $R$. Statistical errors are smaller.
\label{exponents}}

\end{figure}


\begin{references}
\bibitem{oopsI}  We goofed in reference\cite{hysterII}, in constructing
a $O(\epsilon^2)$ diagram for our model which was not present in the
equilibrium models.  The diagram (and all others which differ) turned
out to be irrelevant.
\end{references}
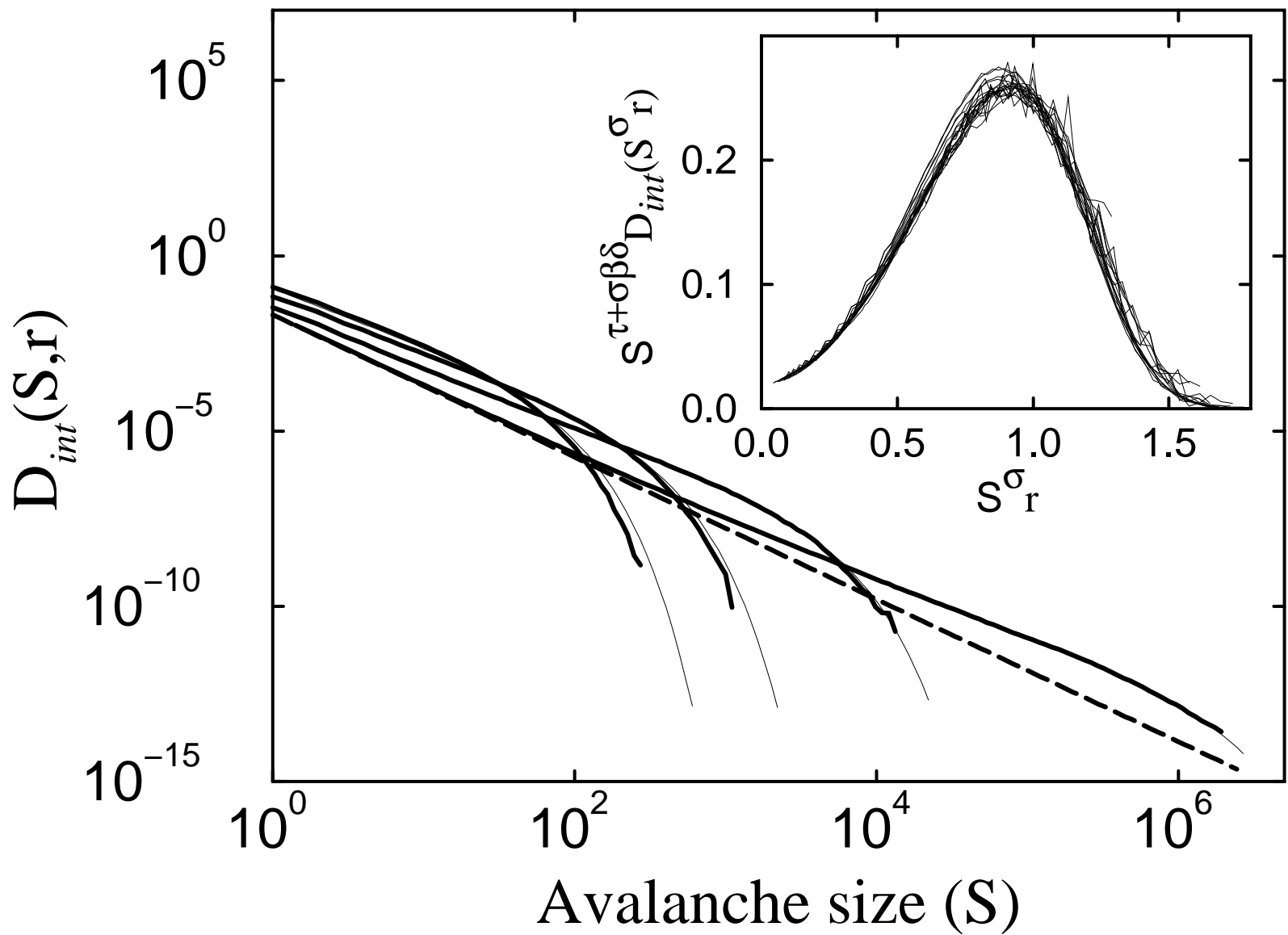
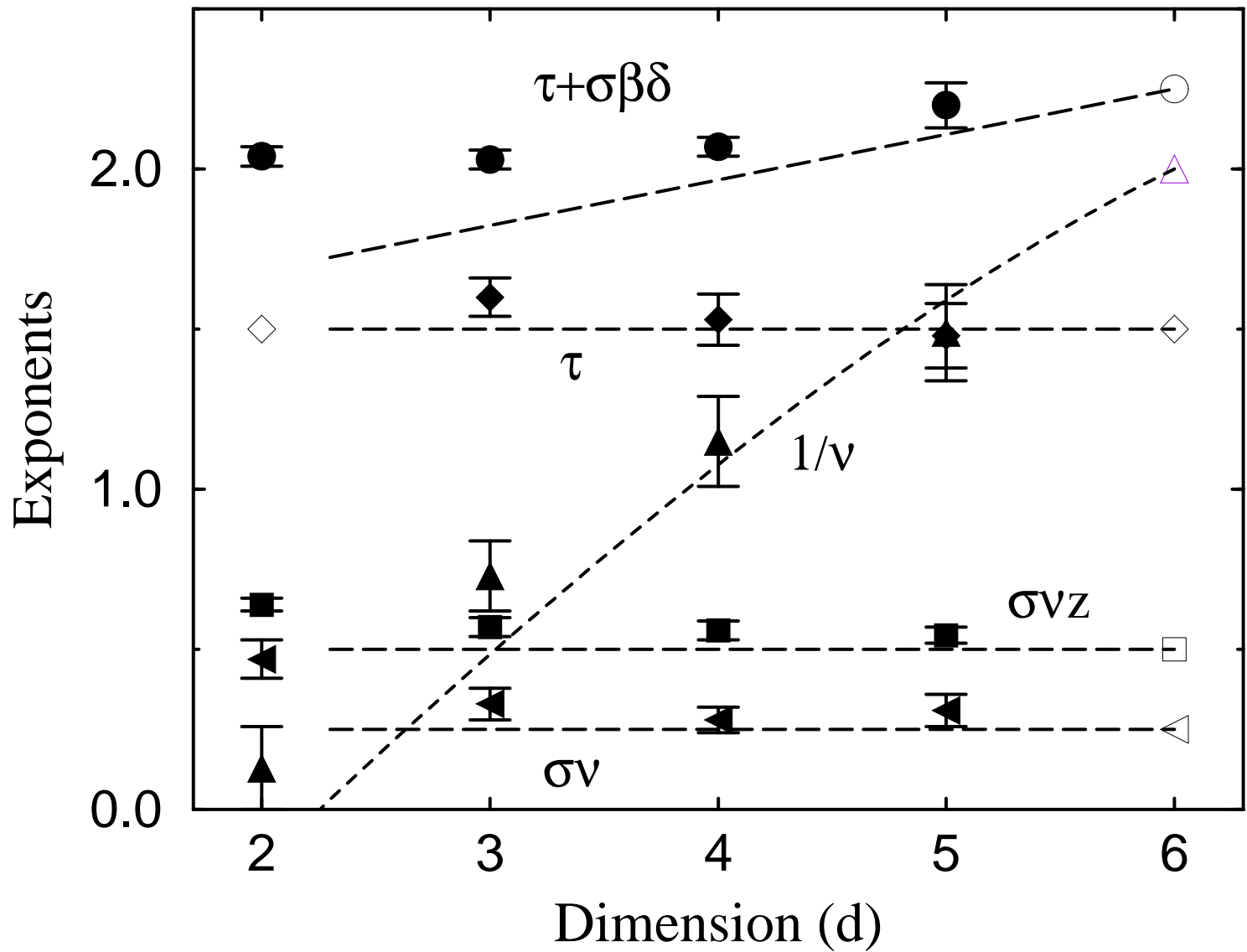
\end{document}